\documentclass[12pt]{article}
\usepackage{amsmath,amssymb}

\newcommand{\rd}[1]{\mathop{\mathrm{d}#1}}

\newcommand{\Tr}{\mathop{\mathrm{Tr}}}

\newcommand{\grad}{\mathop{\mathrm{grad}}\nolimits}
\newcommand{\prob}{\mathop{\mathrm{prob}}\nolimits}
\newcommand{\fract}[2]{{\textstyle\frac{#1}{#2}}}
\newcommand{\bra}[1]{\left\langle #1 \right|}
\newcommand{\ket}[1]{\left| #1 \right\rangle}
\newcommand{\bs}[1]{\boldsymbol{#1}} 
\newcommand{\ts}{\thinspace}
\newcommand{\lips}{\ts.\ts.\ts.\ts} 
\newcommand{\plips}{\lips. }
\newcommand{\E}{\mathcal E}

\let\vec\mathbf 

\input BoxedEPS
\SetRokickiEPSFSpecial  
\HideDisplacementBoxes

\let\Cite\cite

\newcommand{\numeq}[2]{\begin{equation} #2
\label{#1}
\end{equation}}
\newcommand{\refeq}[1]{(\ref{#1})}

\begin{document}
\title{\huge The Depth and Breadth of John Bell's Physics}
\author{R. Jackiw\\
\small\it Center for Theoretical Physics\\
\small\it Massachusetts Institute of Technology,  Cambridge, MA
02139\\[2ex]
 A. Shimony\\
\small\it  Department of Philosophy and Physics (Emeritus)\\
\small\it Boston  University, Boston, MA 02215\\[1ex]
\small Typeset in \LaTeX\ by Martin Stock}
\date{\small MIT-CTP-3141\quad physics/0105046} 

\maketitle
\vspace*{-2pc}

\thispagestyle{plain}

\section{A Brief Biography}\label{s1}

John Stewart Bell (1928--90) was born in Belfast, Northern Ireland, son of
John Bell  and Elizabeth Mary Ann Bell  (n\'ee Brownlee and
always  called Annie), whose families had lived in Northern Ireland for several 
generations. His middle name, Stewart, was a Scottish family name in his
mother's lineage, and until he went to university he was called by that name. The
family's religion was Anglican (Church of Ireland), but no  religious bias prevented
friendships with the Catholic community. The Bell  family was of the working class,
but with a high evaluation of education.  Although only John remained in school
beyond age fourteen, each of his  younger brothers advanced socially, one (who
resumed his studies after  marrying and having children) to become a professor of
electrical engineering  and the other to become a successful businessman.

   By age eleven John, who read intensively at the public library,
announced his intention to become a scientist. He performed very well
on his entrance examination for secondary education, but the family
could only afford to send him to the Belfast Technical High School,
which -- usefully for his later career -- provided both academic and
practical courses. 

   Finishing the Technical School in 1944 at age sixteen, Bell found a job as
a junior technical assistant in the Physics Department of the Queen's
University at Belfast, under the supervision of Prof.~Karl Emeleus and
Dr.~Richard Sloane. They recognized his talent,  loaned him books, and
allowed him to attend first year lectures even though he was not
enrolled as a student. After one year as a technician he became a
student and was awarded modest scholarships, graduating in 1948
with a first class Honours BSc in Experimental Physics. He then stayed
another year, receiving a first class Honours BSc in Mathematical
Physics in 1949. Bell considered himself especially fortunate to have
as his primary teacher in mathematical physics Peter Paul Ewald, a
refugee from Nazi Germany, who was one of the pioneers of X-ray
crystallography. Although Bell received good training in basic physics,
he was not happy with the explanations of quantum mechanics, for
reasons that will given in some detail in Sect.~2. In general, he was
an extraordinarily able and assiduous student, showing
inquisitiveness, persistence, and independence of mind that
foreshadowed the great investigations of his later years. 

From the middle of 1949 until 1960 John Bell worked in England, mainly at 
the UK Atomic Energy Research Establishment (A.E.R.E.) at Harwell. He first 
worked there for about two months on a reactor problem, then went to Malvern 
late in 1949 to join the accelerator group, but returned to Harwell in 1951 to
continue work  on accelerators and to do research on nuclear physics. At Harwell in
1949  John Bell first met his future wife Mary Ross. By that time she had been 
working on reactors for several years. Later, in 1950, she also went to 
Malvern to join the accelerator group. This was her second stay in Malvern, 
as she had been called up during World War II to work on radar at T.R.E. 
Malvern. When the war ended, she returned to Glasgow to take her doctorate in 
physics and mathematics and then went to Harwell early in 1947. They were 
married in 1954 and pursued careers together, often collaborating in 
research, first at Harwell and later at CERN until his death in 1990. There 
is a very moving tribute to her in the last paragraph of the preface to John 
Bell's \emph{Speakable and Unspeakable in Quantum Mechanics}~\cite{n8}, the
collection of his papers on that subject:  ``In  the individual papers I have thanked
many colleagues for their help. But here  I renew very especially my warm thanks
to Mary Bell. When I look through  these papers again I see her everywhere."

John Bell was given a paid leave of absence during 1953  and 1954 to do
graduate studies in the Physics
Department at the University of Birmingham, under the direction of
Paul Matthews and Rudolf Peierls, completing his doctoral thesis in
1956 after returning to Harwell. At Birmingham Bell quickly acquired
a knowledge of current theoretical physics, including quantum field
theory, and his doctoral thesis included a proof of the PTC theorem, on
which he lost priority because L\"uders published a proof of the
theorem before his own paper\cite{n1} appeared in the
\emph{Proceedings of the Royal Society}. As a member of Tony
Skyrme's nuclear group at Harwell he made a number of powerful
applications of field theory and theory of symmetries to nuclear
problems~\cite{r6,r7,r8}.

   In the late fifties, John and Mary Bell were troubled that Harwell was
becoming less devoted to research in fundamental physics, and they
began to look elsewhere for a locus of work. In 1960 they resigned
their tenured positions and took nontenured positions at CERN, he in
the Theory Division and she in the Accelerator Research Group. They
lived in the city of Geneva. Bell's working hours at CERN were devoted
primarily to theoretical particle physics.  He facetiously listed his specialty in an
official CERN document as ``quantum engineering.'' Some of his leisure time was
devoted to the foundations of quantum mechanics, which he called his ``hobby''. A
year of leave in 1963--4,  spent at the Stanford Linear Accelerator
Center, the University of Wisconsin, and Brandeis University, gave him
time to write his two pioneering articles~\cite{n9,n10} on the foundations of
quantum mechanics. Ref.~\cite{n9} demonstrated a conflict concerning
observable quantities between the family of local hidden variables
theories and quantum mechanics. This famous result became known as
``Bell's Theorem'' (see Sect.~2 below).

    On his return to CERN, Bell continued his studies of the foundations
of quantum mechanics privately, but his professional work was in
particle physics, mostly in the framework of the then popular scheme
of current algebra. Within these current algebra investigations Bell
established his famous result on the anomalous nonconservation of
the axial vector current, based on the anomalous breaking of chiral
symmetry~\cite{r26} (see Sect.~3 below).

   Recognition of John Bell as a great scientist came in several stages. In
the early fifties his contributions at Malvern to accelerator design,
based on his expertise in particle dynamics and electromagnetic
theory, were highly regarded;  at age 25 he was a consultant when
CERN was being established. From the time of Bell's graduate studies at
Birmingham  he acquired a high reputation among particle theorists,
which grew and spread worldwide after he joined the Theory Group at
CERN. There was, however,  relatively little attention to his work on
the foundations of quantum mechanics until  experiments (e.g.,
\cite{n11}--\cite{n17}) inspired by Bell's Theorem were reported. Most
of the experimental results were conservative, in that they strongly
supported quantum mechanics and disconfirmed the implications of
the local hidden variables theories, but whatever the results might
have been the experiments surprised the physics community by
showing that hypotheses about  hidden variables and locality,
generally considered to be philosophical  since the famous
Einstein-Bohr debate \cite{n18,n19},  were amenable to empirical
investigation.  ``Bell's Theorem''  became a topic in the guide to authors
of the American Institute of Physics, and interest in the Theorem
spread to philosophers and  (unfortunately with frequent exaggeration
and distortion) to the general public. 

   Bell was honored fairly early by election to the Royal Society in 1972,
proposed by Paul Matthews and Rudolf Peierls. In the eighties many
more honors came to him:   the Reality Foundation Prize (shared with
J.F. Clauser, 1982), Honorary Fellow of the American Academy of Arts
and Sciences (1987), Dirac Medal of the Institute of Physics (1988),
honorary doctorates at the Queen's University of  Belfast (1988) and at
Trinity College in Dublin (1988), the Dannie Heineman Prize for
Mathematical Physics (1989), and the Hughes Medal of the Royal
Society (1989). His contributions to the foundations of quantum
mechanics were regularly cited along with his discoveries in main line
physics. These honors were enhanced by a universal recognition that
John Bell was a man whose character equalled his intellect.

   John Bell died suddenly of a stroke on October 1, 1990. He was still
at the height of his powers just before he was stricken, and the loss to
science was undoubtedly very great. The world also lost a man of
exemplary dedication, integrity, courage, modesty, generosity, and
humanity.

\section{Foundations of Quantum Mechanics}\label{s2}

\subsection{Early Thoughts}

    Even as an undergraduate Bell was seriously troubled by the
foundations of quantum mechanics and unconvinced of the dominant
opinion at his University and in the physics profession generally.
Information about this period can be found in \emph{Quantum
Profiles} \cite{n20} of Jeremy Bernstein, who interviewed Bell, and in
biographical articles of Andrew Whitaker \cite{n21}, who obtained
information from Lesley Kerr, a friend of Bell one year behind him in
the Physics Department at Queen's. Quantum mechanics was taught in
the context of Dr. Sloane's course on atomic spectra, with emphasis on
practical applications rather than conceptual foundations. Bell
questioned Sloane skeptically  and indeed aggressively, and was
dissatisfied not only with Sloane's exposition but with what he learned
generally of the Copenhagen point of view.  The argument with Sloane

\begin{quote}
\lips had convinced Bell of the central importance of the
concept of measurement in any 	meaningful discussion of quantum
theory. It was necessary to identify this as a 	central element, in
order to be able to criticise the fact by pointing out the 	illegitimacy of
using the idea of measurement as a primary term\plips

	   At the time the aspect of this that most concerned Bell was the
divide or 	`Heisenberg cut' between quantum and classical. He fully
accepted Bohr's  acknowledgment that the apparatus must be treated
classically, but definitely did not 	accept his arguments about the
relation between quantum and classical. Actually it 	would be true to
say that he did not recognise where Bohr's actual argument was 
supposed to be, though he knew that Bohr claimed to have solved the
problem. If 	possible Bell wished to eliminate the distinction between
quantum and classical 	altogether, and the most obvious way to do
attempt to do that was by hidden 	variables.\quad\it \cite{n22}, Sect.~5 
\end{quote}

\subsection{Ideas on Measurement and Physical Reality:  ``Beables''}

   The discontent concerning the Copenhagen Interpretation -- and of
quantum mechanics itself, if it is taken to be a complete theory  --
expressed in these reconstructions of Bell's somewhat confused early
ideas persisted throughout his career, but was sharpened, clarified,
and linked to a variety of constructive programs. He felt that Bohr and
Heisenberg were profoundly wrong in giving observation a
fundamental role in physics, thereby letting mind and subjectivity
permeate or even replace the stuff of physics, however much they
obscured that they were doing this.  Observation is indispensable to
the process of obtaining knowledge about the physical world, but  Bell
always maintained that what  is there to be known has an objective
status independently of being observed. Here is a sample of
statements to this effect from publications over a period of a quarter
of a century:

\begin{itemize}
	\item[1967:]  It is easy to imagine a state vector for the whole
universe, quietly 
	pursuing its linear evolution throughout all of time and containing
somehow all 	possible worlds.But the usual interpretive axioms of
quantum mechanics come into 	play only when the system interacts
with something else, is `observed'. For the 	universe there is nothing
else, and quantum mechanics in its traditional form has 	simply
nothing else to say. It gives no way of, indeed no meaning of, picking
out 	from the wave of possibility the single unique thread of
history\plips  In any case it 	seems that the 	quantum mechanical
description will be superseded. In this it is like 	all theories made by
man. But to an unusual extent its ultimate fate is apparent in its 
internal structure. It carries in itself the seeds of its own
destruction.\quad{\it\cite{n23}} 

 	\item[1973:] It is interesting to speculate on the possibility that a
future theory will not be 	  	ambiguous and approximate. Such a
theory would not be fundamentally about 	`measurements', for that
would again imply incompleteness of the system and 	unanalyzed
interventions from outside. Rather it should again be possible to say
of 	a system not that such and such may be \emph{observed}  to be so
but that such and such \emph{be}  so.\quad{\it\cite{n24}}
\goodbreak

	\item[1975:] {\bf The theory of local beables.} This is a pretentious
name for a theory  that 	hardly exists otherwise, but which ought to
exist\plips The terminology, \emph{be-}able as against
\emph{observ-}able, is not designed to frighten with metaphysic
those 	dedicated to realphysic. It is chosen rather to help in making
explicit some notions  already implicit in, and basic to, ordinary
quantum theory\plips `Observables' must be \emph{made}  somehow,
out of beables. The theory of local beables should contain, and 	give
precise physical meaning to, the algebra of local
observables.\quad{\it\cite{n25}}

	\item[1981:]  Where is the `measurer' to be
found?\thinspace\ldots\thinspace  If the theory is to apply to
anything 	but idealized laboratory operations, are we not obliged to
admit that more or less 	`measurement-like' processes are going on
more or less all the time more or less 	everywhere.\quad{\it\cite{n26},
Sect.~1}  

	\item[1990:]  But experiment is a tool. The aim remains to understand
the world. To restrict 	quantum mechanics to be exclusively about piddling
laboratory operations is to betray the great 
enterprise.\quad{\it\cite{n27}}

\end{itemize}

\subsection{Logical Investigations on Hidden Variables Theories}

      As noted above in the quotation from Whitaker's memoir \cite{n22}, Bell was
already attracted in his undergraduate years by the idea of a hidden
variables theory supplementing the description of a physical system
given by quantum mechanics. (He objected later to the term ``hidden
variables'', which suggests inaccessibility to any experimental probing,
and preferred ``beables'', as indicated in one of the quotations above.)
Even without knowing what these hidden variables are in detail, one
might  construct models of them, which would illuminate some of the
conceptual difficulties of quantum mechanics. For instance, fixing
hidden variables might  in principle determine the values of all the
observable quantities recognized by quantum mechanics, though the
chaotic behavior of the physical world at the level of the hidden
variables -- analogous to thermal chaos in kinetic theory of gases --
would preclude experimental control, thereby accounting for
Heisenberg's uncertainties. According to Whitaker, however, Bell's
interest in hidden variables was dampened by reading Max Born's
\emph{Natural Philosophy of Cause and Chance}~\cite{n28}, which
maintained in general terms the difficulty of constructing a theory of
this kind that would recover the known quantum statistics, and in
addition stated that von~Neu\-mann \cite{n29} had given a rigorous
mathematical proof for the impossibility of such a construction. Not
knowing German, Bell could not assess the validity of von~Neu\-mann's
proof, but was inclined for a while to believe Born. 

   We do not seem to have information about the incubation of Bell's
ideas on quantum mechanics between from the time he left Queen's
University until 1952, but in that year he developed them radically.
In the Acknowledgments of Bell's pioneering paper in \emph{Reviews
of Modern Physics}, ``On the Problem of Hidden Variables in Quantum
Mechanics''~\cite{n10} he says, ``The first ideas of this paper were conceived
in 1952. I warmly thank Dr.~F. Mandl for intensive discussion at that
time.'' Mandl was a refugee from Germany and was able to tell him the
content of von~Neu\-mann's proof, as yet untranslated. Furthermore,
David Bohm's ``A Suggested Interpretation of the Quantum Theory in
Terms of `Hidden' Variables''~\cite{n30} was published, providing an instance
of just the sort of model that von~Neu\-mann's theorem allegedly
showed to be impossible. From his reading and his discussions with
Mandl it became clear to Bell that von~Neu\-mann's theorem, though
mathematically correct, was physically weak. One of von~Neu\-mann's
premisses concerned expectation values $\E(A)$, $\E(B)$, etc.\ of
the observables $A$, $B$, etc.\ recognized by quantum mechanics. He
required that in any ensemble of systems of the specified kind, even
an ensemble determined by fixing the hypothetical hidden variables
-- which would be dispersion-free since each observable would be
given a definite value by the fixed hidden variables -- the expectation
values would be additive:
\numeq{e1}{
                                                  \E(A + B) =\E(A) +\E(B)      
 }                          
where A and B are observables of the system quantum
mechanically represented by self-adjoint operators, \emph{not
necessarily commuting}. Bell's analysis, not published until 1966,  is devastating: 
\begin{quote}
The essential assumption can be criticized as follows. At
first sight the required 	additivity of expectation values seems very
reasonable, and it is rather the nonadditivity of allowed values
(eigenvalues) which requires explanation.  Of course	the explanation
is well known: A measurement of a sum of noncommuting 	observables
cannot be made by combining trivially the results of separate 
observations on the two terms -- it requires a quite different
experiment. For example the measurement of $\sigma_x$  for a
magnetic particle might be made with a 	suitably oriented
Stern-Gerlach magnet. The measurement of $\sigma_y$  would require
a  different orientation, and of $(\sigma_x + \sigma_y)$ a third and
different orientation. But this  explanation of the nonadditivity of
allowed values also established the nontriviality 	of the additivity of
expectation values. The latter is a quite peculiar property of 	quantum
mechanical states, not to be expected \emph{a~priori}. There is no
reason to  demand it individually\vadjust{\newpage} of the hypothetical dispersion
free states, whose function it 	is to reproduce the \emph{measurable} 
peculiarities of quantum mechanics \emph{when averaged
over}.\nobreak\quad{\it\cite{n10}, Sect.~3}
\end{quote}

   In addition to criticizing von~Neu\-mann's premisses, Bell constructed a
family of hidden variables models for a spin-1/2 particle in each of
which definite values are consistently assigned to every spin
observable, and the additivity of expectation values holds for
observables represented by commuting operators. Each such model
can be taken as a hidden variable $\lambda$ in  a space $\Lambda$ of
hidden variables,  $\lambda(A)$ being the value assigned by this
hidden variable to the observable~$A$.  Bell showed, furthermore, that 
every quantum mechanical state $\phi$ of the spin-1/2 system can be
constructed by averaging over $\Lambda$ with an appropriate
probability measure $\rho$ (independent of the observables):
\numeq{e2}{
                         \bra\phi   A \ket \phi   =
\int_\Lambda A(\lambda) \rd\rho\ .        
   }                                

Bell's construction is quite simple, and it will be seen momentarily that
the dimension two of the Hilbert space associated with the spin-1/2
system  is crucial. Bell notes that hidden variables model differs
conceptually from Bohm's by using only the algebraic structure of the
observables of the spin-1/2 system, without any consideration of the
apparatus used for measuring.

   An important part of Ref.~\cite{n10} was inspired by a theorem of Andrew
Gleason \cite{n32}, which was not published until 1957 and not known by
Bell until conversations with  Josef Jauch after  moving to Geneva.
Gleason did not directly address the question of hidden variables, but
considered which probability measures are definable on the lattice of
projection operators  (which is isomorphic to the lattice of closed linear
subspaces) on a Hilbert space.  The conditions on a probability
measure $m$ are
\begin{enumerate}

\item   $m(Q)$ is a nonnegative real number for any projection
operator $Q$ on 
		     the Hilbert space; 

\item $m(I) = 1$,  where $I$ is the identity operator;

\item  If $\{Q_i\}$ is a finite or countably infinite set of mutually
orthogonal projection operators, then  $m(\sum Q_i)  = \sum m(Q_i)$.
\end{enumerate}
 Gleason's theorem asserts that if the Hilbert space has dimension
three or greater then a probability measure must be determined
either by a quantum mechanical pure state,
\numeq{e3}{                       
                                       m(Q) =  \bra\phi   Q \ket \phi       
}
 for some vector $\phi$ in the Hilbert space with norm unity, or else by
a convex combination of measures thus determined; equivalently, $m$
must be determined in the standard quantum mechanical manner, 
\numeq{e4}{
                                    m(Q) = \Tr(QW)                                
}
where $W$ is a statistical operator, i.e., a positive self-adjoint operator
of trace unity.

   One easily sees that  $\Tr(QW)$ cannot have values restricted to 0 or
1 for all projection operators $Q$, from which in turn it follows that the
state represented by $W$ cannot be dis\-persion-free for each
observable, that is, each self-adjoint operator. Consequently, a
corollary of Gleason's theorem is that no system associated quantum
mechanically with a Hilbert space of dimension three or greater can
admit a dispersion-free state;  or in von~Neu\-mann's locution, there can
be no dispersion-free ensemble of such systems. And of course, since
the set of dispersion-free states is empty, one cannot  recover a
quantum mechanical state $\phi$ by averaging appropriately over
them.   Note that the hidden variables model constructed by Bell
earlier in Ref.~\cite{n10} does not conflict with Gleason's theorem and its
corollary,  since the Hilbert space in that case has dimension two.

   The proof given by Gleason of his theorem is intricate; Bell said (at a
conference) that he knew he had either to read Gleason's proof or to
construct his own proof of the corollary that interested him, and the
latter alternative was obviously easier. And this he did, elegantly and
simply, in Sect.~5 of Ref.~\cite{n10}. An independent proof of the corollary
was published the following year by Simon Kochen and Ernest
Specker~\cite{n33}. 

    After this accomplishment, the end of Bell's Ref.~\cite{n10} is a
remarkable surprise. Having exhibited that the hidden variables
program, construed as it had been in the past, is mathematically
impossible except in the trivial case of dimension two, Bell
resuscitated the program by a well-motivated relaxation of conditions.
He slyly remarked, ``That so much follows from such apparently
innocent assumptions leads us to question their innocence.'' What is
not innocent, he pointed out, is the tacit assumption that when the
hidden variable $\lambda$ is specified, ``the measurement of an observable
must yield the same value independently of what other measurements
may be made simultaneously.'' If observable $A$ and $B$ commute,
then quantum mechanics requires that in principle both can be
measured simultaneously, because there exists an observable C such
that $A$ can be expressed as a function $A(C)$ and $B$ as a function
$B(C)$. But $A$ can commute with two observables $B$ and $B'$ that 
are noncommuting, and then a $C$ of which $A$ and $B$ are functions
necessarily differs from a $C'$ of which both $A$ and $B'$ are
functions. Thus a different procedure is required to measure A along
with $B$ from the one used to measure $A$ along with $B'$. This
observation opens up a new family of hidden variables theories, which
have come to be called ``contextual'', in which the value of an
observable $A$ is a function not only of the hidden variable $\lambda$
but also the context of measurement. Bohm's model \cite{n30} was an
instance of this idea,  but he did not present the idea explicitly and
generally. By a judo-like maneuver Bell elicits unwitting cooperation
in the resuscitation of the hidden variables program from Niels Bohr, a
dedicated opponent of it, by citing Bohr's thesis of the ``impossibility
of any sharp distinction between the behavior of atomic objects and
the interaction with the measuring instruments which serve to define
the conditions under which the phenomena appear''~\cite{n34}.

\subsection{Bell's Theorem}

   The most influential of Bell's papers on the foundations of quantum
mechanics was ``On the Einstein-Podolsky-Rosen Paradox''~\cite{n9}.  Published in
1964,  its content presupposes \cite{n10}, which was published two years later.
The explanation for this chronological peculiarity is the misplacement of \cite{n10}
in the offices of \emph{Reviews of Modern Physics}. Ref.~\cite{n9} is the pioneering
presentation of what has come to be called ``Bell's Theorem,'' which (roughly)
asserts that
\emph{no hidden-variables theory that satisfies a certain locality
condition can reproduce all the predictions of quantum
mechanics}. The name ``Bell's Theorem'' actually applies to a
family of theorems with the general character just stated, but we shall
review only three variants that Bell himself proved. In all of Bell's
variants a space of \emph{complete specifications -- the hidden
variables} -- of the system is envisaged (not the case in other variants,
notably the controversial ones of Henry Stapp \cite{n35, n36} and other
expositions), and the results are independent of the detailed character
of the hidden variables. What is important is (i)~the way in which
observable quantities depend upon the hidden variables, and (ii)~the
fact that the space of hidden variables admits probability
distributions. In Ref.~\cite{n9} the hidden variable is taken to be a single
continuous parameter, and the system studied for the purpose of
proving the Theorem consists of two well-separated particles 1 and 2.
A family of quantities parameterized by $\vec a$ is measurable on 1,
each, for simplicity, assumed to have  only two possible outcomes $+1$
and $-1$;  likewise, a family of quantities parameterized by $\vec b$ is
measurable on~2, with the same assumption regarding outcomes. The
outcome of a measurement on 1 is a function
\begin{subequations}\label{e5}
\numeq{5a}{
					A(\vec a, \lambda) = \pm1 
}
regardless of what quantity is measured on 2;  and likewise the
outcome of  a measurement on 2 is a function\qquad
\numeq{5b}{
				        B(\vec b, \lambda)  = \pm1}                                       
\end{subequations}
regardless of what quantity is measured on 1. Since the outcomes are definite when the
quantity specified by $\vec a $ (respectively $\vec b$) is given along with 
$\lambda$, the hidden variables theory is \emph{deterministic}. It may appear that
Bell has neglected the possibility of contextuality, which he introduced in
Ref.~\cite{n10} to resuscitate the hidden variables program, but that impression is
due to the notation. As a matter of fact, two different parameters $\vec a$ and
$\vec a '$ for particle 1 could refer to the same measurable quantity  measured
along with different contexts concerning particle 1, though with no reference to
what is measured on 2;  and likewise for $\vec b$ and
$\vec b '$. Contextuality is tacitly allowed in this version of Bell's theorem,
but it is restricted by a {\it  locality requirement}, which is vital: {\it the result
A does not depend upon what is measured on} 2, {\it and the result B does
not depend upon what is measured on} 1.  Bell justifies this
requirement with a quotation from Einstein.
	\begin{quote}
But on one supposition we should, in my opinion, hold absolutely
fast:  the real 	factual situation of the system  $S_2$ is independent of
what is done with the system  $S_1$, 	which is spatially separated from
the former.\quad\it\cite{n9}, footnote~2 
 \end{quote}
In Ref.~\cite{n9} the well-separated particles 1 and 2 are assumed to have spin-1/2  
and to be in the quantum mechanical singlet state. If $\vec a$ and $\vec b$ are vectors in
ordinary three-space,  then $\bs\sigma _1\cdot \vec a$
is the quantity parameterized by $\vec a$,  and $\bs\sigma_2\cdot \vec b$ is the
quantity parameterized by~$\vec b$. The anti-correlation of spins of particles in the
singlet state guarantees that if  $\vec a$ and  $\vec b$ are equal then the outcomes of
the measurements of the corresponding quantities of 1 and 2 are
opposite, or equivalently, the product of their measurement outcomes
is $-1$.  For the hidden variables theory to agree with this quantum
mechanical prediction there must be a probability distribution $\rho$ over
the space of hidden variables (normalized to unity) such that 
\numeq{e6}{
                                  \int \rd \lambda \rho(\lambda)A(\vec
a,\lambda)B(\vec a,\lambda) =-1. }                                     
But this is possible only if
\numeq{e7}{
                                           A(\vec a,\lambda) =-B(\vec a, \lambda)  
}                                                     
for all but a set of measure zero in the space of hidden variables.
   All the concepts and assumptions of the theorem in Ref.~\cite{n9} have now
been introduced, and the remainder of Bell's work is mathematics,
which we shall condense.  Using Eqs. \refeq{e5}, \refeq{e6}, and 
\refeq{e7} he easily proves the inequality
\numeq{e8}{
    1 + P(\vec b, \vec c)   \geq  \bigl|  P(\vec a, \vec b) -
 P( \vec a, \vec c)\bigr|   }                            
where 
 \numeq{e9}{                                             
 P(\vec a, \vec b) = \int \rd \lambda \rho(\lambda) A(\vec a,\lambda)
B(\vec b,\lambda)
   }                           
(that is, the expectation value using the probability distribution $\rho$ of the
product of the outcomes of measurement of the spin components parameterized by
$\vec a$ and $\vec b$). Inequality \refeq{e8} is the prototype of a family of
inequalities that are now collectively called ``Bell's Inequalities.'' It is a well-known
consequence of the quantum mechanical singlet state that the expectation value of
the product of an arbitrary spin component of particle 1 and an arbitrary spin
component of particle 2 satisfies 
\numeq{e10}{
    \bigl\langle\bs\sigma_1\cdot \vec a\  \bs\sigma_2\cdot\vec b\bigr\rangle =
-\vec a
\cdot
\vec b\ .}
The discrepancy between the
hidden variables inequality \refeq{e8} and the quantum mechanical
expectation value \refeq{e10} is immediate, thus proving the theorem
of Ref.~\cite{n9}.

   There are  important additional points in this paper. The
title, ``On the Einstein-Podolsky-Rosen [EPR] Paradox'', signals the
importance of EPR's paper \cite{n18} in focusing on a pair of well-separated
particles that has been properly prepared to ensure strict correlations
between certain of their observable quantities. Bell recognizes in
Sect.~2 that the deterministic character of the hidden variables
theory that EPR recommend is not an additional assumption, but is the
consequence of EPR's assumptions of locality, exact agreement with the
quantum mechanical predictions of strict correlation, and their well-known 
sufficient condition for the existence of an element of physical
reality (but his acknowledgment of the third assumption is tacit rather
than explicit). Next, in recognition of the inaccuracies of actual
measurement,  Bell asks in Sect.~4 of  Ref.~\cite{n9}  whether a deterministic
local hidden variables theory can agree {\it approximately}  with the
quantum mechanical predictions, and he provides a negative answer
by deriving a generalization of inequality \refeq{e8}. He also briefly
mentions generalizations of his theorem to pairs of systems each
associated with a Hilbert space of dimension greater than two and
having observables with more than two possible values. Finally, and
most remarkably, in Sect.~6 he mentions the possibility that quantum
mechanics is of limited validity, applying only when ``the settings of
the instruments are made sufficiently in advance to allow them to
reach some mutual rapport by exchange of signals with velocity less
than or equal to that of light.'' If so, the locality requirement assumed
in the theorem would not be physically justified and would not be a
consequence of the locality of special relativity theory. In order to
check this conjecture, ``experiments of the the type proposed by Bohm
and Aharonov\lips  in which the settings are changed during the
flight of the particles, are crucial'' \cite{n38}. Thus, in this pioneering
paper of 1964 Bell already suggests the desirability of an experiment
like that carried out by Aspect, Dalibard, and Roger in 1982~\cite{n17}. 

   In 1971 Bell \cite{n39} published a new version of his Theorem, which is
an improvement over the first in two respects.  (a)~It dispenses with
reliance upon  quantum mechanics, as in Eq.~\refeq{e2} above, for the
purpose of deriving a crucial inequality.  In so doing it follows a paper
by Clauser {\it et al}.~\cite{n40} of 1969. If the resulting inequality 
conflicts with experiments, one can conclude not only the falsity of
the conjunction of quantum mechanics with the assumptions of the
local hidden variables theory  but the falsity of those assumptions
themselves without conjoining quantum mechanics. (b)~It weakens the 
assumption \refeq{e5} that the hidden variable deterministically
governs the outcome of each observable quantity of particles 1 and 2,
asserting instead that $\lambda$ fixes the expectation value of the
outcome of measuring the quantities of 1 parameterized by $\vec a$,
or equivalently the probability of each possible measurement outcome,
and likewise for the quantities of 2 parameterized by $\vec b$.
\emph{The locality requirement is that the expectation value of the
former is independent of\/  $\vec b$ and conversely.}   In Sect.~4 Bell
says that the failure of deterministic equations like \refeq{e5} above
may be due to hidden variables of the instruments, but in Footnote 10
he adds that this failure may be due to a more fundamental
indeterminism, persisting even when the hidden variables both of the
particles and the instruments are specified. The character of the
indeterminism does not affect the mathematics which leads to the
inequality
\numeq{e11}{
    \bigl| P(a,b) - P(a,b')\bigr| + \bigl|P(a',b') +P(a',b)\bigr|  \leq 2\ .
   }
 It is easy to show the inconsistency of  Inequality \refeq{e11} with
the ideal quantum prediction expressed by Eq.~\refeq{e10} and -- obviously
valuable in experimental investigations -- even with approximations to
the ideal prediction. In view of the assumptions upon which Inequality
\refeq{e11} is based, its disconfirmation by experiment would throw doubt
upon the entire family of deterministic and nondeterministic local
hidden variables theories.

Bell's third variant of his Theorem appeared in a somewhat rough
form in ``The theory of local beables'' \cite{n25}, but a better
presentation, meeting some criticisms, appeared in ``La Nouvelle
Cuisine'' \cite{n41}, written too late to be included in the collection
\emph{Speakable and Unspeakable in Quantum Mechanics}. What is
interesting about this variant is a greater explicitness about the
locality requirement than in its predecessors. He supposes that the
measurements of the quantities of quantities parameterized
respectively by a and b are performed in regions~1 and~2   with
space-like separation. (For the careful use of special relativity theory
this description of the ideal experimental arrangement is clearly more
precise than ``well-separated particles''.)  Region~3 is bounded by two
nonintersecting space-like surfaces, both cutting through the
backward lightcones of both 1 and 2 in such a way that the
intersections of the later surface with the two backward lightcones are
nonoverlapping, while the intersections of the earlier surface with the
two backward lightcones overlap. In region~3, let $c$ stand for any set of
physical quantities describing the experimental arrangement and
recognized by quantum mechanics in that part of region 3 that blocks
the two backward lightcones and $\lambda$ be all other variables
needed in addition to $c$ to describe completely this part of region~3.
The probability of joint outcomes $A$ and $B$ of the two
measurements reasonably depends only upon $a$, $b$, $c$, and
$\lambda$ and hence can be written as
$ \bigl\{ A,B\bigm| a,b,c,\lambda\bigr\}$, which by standard
probability theory can be factorized into  
\numeq{e12}{                                    
      \bigl\{ A,B\bigm| a,b,c,\lambda\bigr\} = \bigl\{
A\bigm| B,  a,b,c,\lambda\bigr\}  \bigl\{
B\bigm| a,b,c,\lambda\bigr\} \ .}                           Then, Bell reasons, 
\begin{quote}       Invoking local causality, and the assumed
completeness of $c$ and $\lambda$ in the relevant parts of region 3,
we declare redundant certain of the conditional variables in the last
expression, because they are at space-like separation from the result
in question. Then we have
\end{quote}    
\numeq{e13}{
  \bigl\{ A,B\bigm| a,b,c,\lambda\bigr\} =  \bigl\{ A\bigm| 
a,c,\lambda\bigr\}
  \bigl\{ B\bigm| b,c,\lambda\bigr\} \ .} 
Bell remarks that Eq. \refeq{e13} ``exhibits $A$ and $B$ as having no
dependence on one another, nor on the settings of the remote
polarizers ($b$ and $a$ respectively), but only on the local polarizers
($a$ and $b$ respectively) and on the past causes, $c$ and $\lambda$.''  
Eq.~\refeq{e13} is essentially the locality requirement used in
Ref.~\cite{n39}, but here he does not take it as the starting point of his
analysis but rather as a consequence of two quite different locality
assumptions used in passing from Eq. \refeq{e12} to Eq. \refeq{e13}.
(There is similar reasoning in Sects. 2 and 4 of  Ref.~\cite{n25} and in a
paper by J. Jarrett \cite{n42}.)  Once Eq.
\refeq{e13} is derived, and it is further reasoned that the probability
distribution for the hidden variable $\lambda$ is independent of the
freely chosen variables $a$ and $b$ and therefore dependent only upon
$c$, one obtains a factorizability condition for expectation values: 
\numeq{e14}{
  \E(a,b,c) =\sum_\lambda \sum_{A,B} AB\bigl\{
A\bigm|  a,c,\lambda\bigr\} \bigl\{ B\bigm|  b,c,\lambda\bigr\}  \bigl\{\lambda
\bigm| c \bigr\} .                } Inequality (11) can then be derived
essentially as in Refs. \cite{n39} and
\cite{n40}, and Bell omits the details.

\subsection{Is There Nonlocality in Nature?}\label{ss:E}

    In Sect.~6.7 of \cite{n41} Bell offers the following formulation of the
principle of local causality:  ``The direct causes (and effects) of events
are near by, and even the indirect causes (and effects) are no further
away than permitted by the velocity of light.'' He then reviews Bell's
Theorem and the experiments which it inspired to investigate whether
the principle of local causality holds in quantum mechanics and in
nature. He pays attention to attempts to give a positive answer to this
question, but concludes that these are desperate and unconvincing.
Bell's own negative answer led him to  controversial but tentative
proposals, which will be presented below.

   The ideal predictions of quantum mechanics in certain situations violate
Inequality \refeq{e11} -- for instance, when the linear polarization of photon 1 of a
pair of photons is measured in region 1 of Ref.~\cite{n41} and the linear
polarization of photon 2 of the pair is measured in region~2, the pair being
prepared in the polarization state
\numeq{e15}{
  \Psi = \fract {1}{\sqrt 2} \bigl\{\ket h_1 \ket h_2 + \ket v_1 \ket v_2\bigr\}
}                      
  where $\ket h$ is a state of horizontal polarization and $\ket v$ a state of vertical
polarization. Consequently, quantum mechanics must deny one or both of the two
omissions of variables in the transition from Eq.
\refeq{e12} to Eq. \refeq{e13}. As a matter of fact, it is immediate, in view of the
correlations of Eq. \refeq{e10}, that quantum mechanics does not permit the
substitution of $\bigl\{ A\bigm| a,c,\lambda \bigr\}$ for $\bigl\{ A\bigm|
B,a,b,c,\lambda
\bigr\}$ (where $\lambda$ is simply the quantum state $\Psi$ in quantum
mechanics). On the other hand, the substitution of $\bigl\{ B\bigm| b,c,\lambda
\bigr\}$ for $ \bigl\{ B\bigm| a,b,c,\lambda \bigr\}$ has often been shown to be
valid in quantum mechanics;  a sketch of an argument in the case of  relativistic
quantum mechanics -- using the commutativity of Heisenberg operators at
space-time points that are space-like separated --  is given by Bell in Secs. 6.5 and
6.11 of Ref.~\cite{n41}. If this substitution were not allowed, then a superluminal
signal could be transmitted from region 1 to region  2, since dependence of the
probability of outcome $B$ upon the freely chosen variable~$a$ provides
(probabilistically) some information about that choice. Alternatively, the
dependence -- entailed by quantum mechanics -- of the probability of the outcome
$A$ in region 1 upon the outcome $B$ in region 2 does not permit transmission of
information about a choice, because the outcome $B$ is a chance event not chosen
by the experimenter in region 2. In other words, although quantum mechanics
violates Inequality \refeq{e11}, which is a consequence of any local hidden
variables theory, deterministic or nondeterministic, it does so in such a way that the
violation cannot be \emph{used} to provide superluminal communication.
Accordingly, some writers on the problem (included one of us, A.S.) have suggested
that, despite tension between quantum mechanics and special relativity theory,
there is ``peaceful coexistence'' between the two theories. Furthermore, if one
accepts  the auxiliary assumption that the pairs of particles detected in an actual
correlation experiment is a fair sample of the pairs emerging from polarization
analyzers, even though the detectors are quite inefficient -- an assumption that
Bell regards as very reasonable (\cite{n41}, Sect.~6.12) -- then in view of the
overwhelming  favoring of quantum mechanics in correlation experiments there is
a kind of nonlocality in nature, but such as to coexist peacefully with relativistic
locality. 

   Bell is scornful of this optimistic and conciliatory proposal. The
reason is his unwillingness to admit that something complex, special,
and anthropocentric can be a fundamental principle of physics -- the
same reason that is pervasive in  his polemic ``Against
`Measurement'\ts'',  cited in Sect.~2.2 above.
\begin{quote}
	Do we then have to fall back on `no signaling faster than light' as the
expression of 	the fundamental causal structure of contemporary
theoretical physics? That is hard 	for me to accept. For one thing we
have lost the idea that correlations can be 	explained, or at least this
idea awaits reformulation. More importantly, the `no 
signaling\lips'  notion rests on concepts that are
desperately vague, or vaguely  applicable. The assertion that `we
cannot signal faster than light' immediately 	provokes the question:

\centerline{Who do we think \emph{we}  are?}
\noindent
	 \emph{We}  who can make `measurements', \emph{we}  who can manipulate
`external fields', \emph{we} 	who can `signal' at all, even if not faster than
light? Do \emph{we} include chemists, or only 	physicists, plants, or only
animals, pocket calculators, or only mainframe 	computers?
\quad \cite{n41}, Sect.~6.12 
\end{quote}
 This polemic against the fundamental significance of signaling implies that the
failure according to quantum mechanics of the reduction of $\bigl\{ A\bigm|
B,a,b,c,\lambda \bigr\}$ to $\bigl\{ A\bigm| a,c,\lambda  \bigr\}$, in the transition
from Eq. \refeq{e12} to Eq. \refeq{e13} is as serious a violation of local causality as
the failure of the transition $\bigl\{ B\bigm| a,b,c,\lambda \bigr\}$ to $ \bigl\{
B\bigm| b,c, \lambda \bigr\}$ would have been. 

    Among the logical possibilities that Bell tentatively explores in
order to escape from this disagreeable conclusion is to envisage a
theory in which the apparent free will of the experimenters in
choosing the variables a and b is an illusion. But he comments, 
\begin{quote}
Perhaps such a theory could be both locally causal and in agreement
with quantum 	mechanical predictions. However I do not expect to see
a serious theory of this 	kind. I would expect a serious theory to
permit `deterministic chaos', or 	`pseudorandomness', for complicated
subsystems (e.g., computers) which would 	provide sufficiently free
variables for the purpose at hand. But I do not have a 	theorem about
that\plips\quad\it\cite{n41}, Sect.~6.10 
\end{quote}
\goodbreak

     Another logical possibility (mentioned in Ref.~\cite{n41}, Sect.~6.12) after a
preliminary discussion of the concept of entropy) is that ``causal
structure emerges only in something like a `thermodynamic'
approximation, where the notions `measurement' and `external field'
become legitimate approximations.'' But he cannot believe that this is
the whole story, because ``local commutativity  [which is invoked for
the derivations of dispersion relations in quantum field theory] does
not for me have a thermodynamic air about it. It is a challenge now to
couple it with sharp internal concepts, rather than vague external
ones. Perhaps there is already a hint of this in `quantum mechanics
with spontaneous wave function collapse.'\,''  In Sect.~2.6  we shall
discuss some of Bell's conjectures on the quantum mechanical
measurement problem, for which spontaneous wave function collapse
is a recent proposed solution. For the present, we merely note Bell's
entertainment of the idea that there is an intimate connection between
the locality problem and the measurement problem. 

    Another logical possibility considered by Bell is a reversion to the
pre-Einsteinian relativity theory of Fitzgerald, Larmor, Lorentz, and
Poincar\'{e}, in which there is a preferred reference system (an
``aether''), but forces dependent upon velocity relative to the
preferred frame are responsible for Fitzgerald-Lorentz contraction,
hence for the null result of the Michelson-Morley experiment, hence
for the cover-up that prevents the identification of the preferred
frame. In this pre-Einsteinian theory there is no kinematical obstacle to
superluminal causal influences. Consequently, the transition above
from Eq. \refeq{e12} to Eq. \refeq{e13} would be blocked because 
$ \bigl\{ A\bigm| B,a,b,c,\lambda \bigr\}$ could
indeed depend upon the variable $b$, and upon the chance outcome of
the experiment parameterized by $b$,  even though that variable is freely
chosen in a region with space-like separation from the region where
variable~$a$ is freely chosen.  Hence there would be a causal relation
between outcomes $A$ and $B$, even though no light signal could connect
the events. Whether $A$ is the cause and $B$ the effect or conversely
depends upon the time order in the preferred reference system. 
Although this method of permitting a causal connection between two
regions with space-like separation sacrifices the conceptual beauty of
Einstein's special theory of relativity, much and perhaps all of the
experimental consequences of that theory is recovered in a sufficiently
clever reversion to the pre-Einsteinian theory, and Bell's paper ``How
to Teach Special Relativity''~\cite{n43} argues, nonpolemically,  that there are
some pedagogical advantages in this reversion. But the main
motivation for going back to Fitzgerald et al., once the peculiarities of
quantum mechanical correlations have been established
experimentally, is that we would have a space-time theory that
accommodates these peculiarities. Having entertained this solution to
the problem of quantum mechanical nonlocality, however, Bell cannot
endorse it without serious reservations: 
\begin{quote}
As with relativity before
Einstein, there is then a preferred frame in the formulation of the
theory\lips   but it is experimentally indistinguishable\plips  It seems
an eccentric way to make a world.~\quad\it\cite{n44}, last paragraph 
\end{quote}

   We suspect that the passage in Bell's writings that most accurately
expresses his attitude towards the apparent nonlocality of quantum
theory, which he more than any one else brought into sharp focus, is
an unspecific and open conjecture about the physics of the future:  ``It
may be that a real synthesis of quantum and relativity theories
requires not just technical developments but radical conceptual
renewal''~(\cite{n45}, last sentence).

\subsection{Unpromising and Promising Approaches to
 the Measurement Problem}

    The quantum mechanical measurement problem is generated by the
unitarity of the quantum mechanical time-evolution operators
together with the assumption that quantum mechanics governs the
entire physical world. As a consequence, when a physical apparatus
interacting with an object of interest physically indicates (e.g., by a
``pointer reading'' or some generalization thereof) the eigenvalue of an
observable quantity whenever the object is initially in an eigenstate of
that quantity with that eigenvalue, then in the more general case of an
initial object state which is a superposition of eigenstates with
differing eigenvalues the ``pointer'' of the apparatus at the conclusion
of the measurement has an indefinite reading. How then can there be
definite outcomes of measurements, whose probabilities are supposed
to be predicted by the formalism of quantum mechanics?  Bell
comments in some detail on four different attempts to solve this
problems, two of which he considers to be unsuccessful and two at
least promising.
 
\subsubsection*{(i) The Copenhagen Interpretation} 

The most unsuccessful, in Bell's opinion, is historically the most widely
accepted -- the Copenhagen interpretation of quantum mechanics in
one version or another. ``Against `Measurement'\ts"~\cite{n27} is Bell's polemic
against this interpretation, represented by the versions of L.D. Landau
and E.M. Lifshitz~\cite{n46}, K. Gottfried~\cite{n47}, and N.G.
van~Kampen~\cite{n48}, which he regards as sufficiently strong and characteristic
to represent adequately the entire school. Each version makes, explicitly or
implicitly, certain assumptions and approximations that are good ``for all practical
purposes'' -- a phrase which he abbreviates with the biting acronym ``FAPP''.  
Landau and Lifshitz (LL) emphasize the macroscopic constitution of the apparatus
actually used in measurement, in virtue of which it behaves as a classical physical
system. But that is consistent with the quantum mechanics only if
there is a spontaneous transition of the state of the apparatus into an
eigenstate of its ``pointer reading''. But how {\it in detail}  is this transition
-- irreversible according to LL  -- compatible with the universal
principles of quantum mechanics? Bell concludes,
\begin{quote}
The LL formulation, with vaguely defined wave function collapse,
when used 	with good taste and discretion, is adequate FAPP. It
remains that the theory is 	ambiguous in principle, about exactly
when and exactly how the collapse occurs, 	about what is microscopic
and what is macroscopic, what quantum and what 	classical. We are
allowed to ask:  is such ambiguity dictated by experimental facts?
	Or could theoretical physicists do better if they tried harder?
\end{quote}
 
    An essential element in Gottfried's solution to the measurement
problem is first to represent  the final state of the object and
apparatus  (neglecting the interaction of the apparatus with the rest of
the world) by a density matrix $\rho$ in the pointer-reading basis, which
will in general contain off-diagonal terms, and then to retain only the
diagonal elements of $\rho$, yielding the density matrix
\numeq{e16}{
   \hat\rho  = \sum_n \bigl|c_{nn}\bigr|^2 \Psi_n \Psi^*_n\ .
}
 Bell is willing to allow the harmlessness FAPP of
replacing of $\rho$ by $\hat\rho$, because of the ``practical elusiveness,  even
the absence FAPP, of interference between macroscopically different
states''. But he objects strenuously to Gottfried's assertion that the right-hand side
of Eq. \refeq{e16}  endows the coefficients $|c_{nn}|^2$ with an interpretation as
probabilities. 
\begin{quote}
	I am quite puzzled by this. If one were not actually on the lookout for
probabilities, I think 	the obvious interpretation of even $\hat\rho$ would be
that the system is in a state in which 	the various $\Psi$s somehow
coexist\plips This is not at all a probability interpretation, in 	which the
different terms are seen not as coexisting, but as alternatives\plips The
idea 	that the elimination of coherence, in one way or another, implies
the replacement of 	`and' by `or' is a very common one among solvers
of the `measurement problem'. 	It has always puzzled me. 
\end{quote}
 Bell goes
on to argue that in spite of Gottfried's pretense at obtaining
irreversibility out of the reversible exact quantum dynamics by an
approximation, he is modifying the rules of quantum mechanics -- 
surreptitiously and without precision. 

Van~Kampen divides the entire world $W$ into ``system'' $S$,
``apparatus'' $A$, and the rest $R'$, so that $W$ can be written as $S+A+R'  =
S' + R'$, and he cleverly focuses on the $S'/R'$ interface. One treatment
of the interface is rigorously quantum mechanical. The other
treatment takes into account measurements that can actually be done,
and these FAPP show no interference between macroscopically
different states of $S'$. Bell attributes to van Kampen a conjunction of
these points of view with the consequence,
\begin{quote}
	It is $\it as \, if$ the `and' in the superposition had already, $\it before$  any
such  measurements, been replaced by `or'. So the `and' $\it has$  already
been replaced by 	`or'. It is $\it as \, if $ it were so\lips  so it $\it is$  so.
\end{quote}
 Bell
concludes that either by a quantum mechanical theorem or by a
change of theory  van Kampen has reached a conclusion like that
attributed to LL:  that a superposition of macroscopically different
states somehow decays into one of its members. But no theorem is
given, and Bell finds it unlikely that one could be given, because the
only procedure of calculation available for proving such a theorem
would require a further shift of the slippery split between system and
the rest of the world, and the indefiniteness of the split precludes
precise calculation. This negative conclusion points, for Bell,  in the
direction of changes of theory, samples of which will be sketched in
(iii) and (iv) below.

\subsubsection*{(ii) H. Everett's Interpretation}

     There appear to be two reasons why Bell discusses Everett's interpretation at
some length in two papers ~\cite{n49,n26}. The first is positive -- that Bell
appreciates Everett's rejection of a division of the world into a part treated
quantum mechanically and a part treated classically (which is Bell's b\^ete noir in
the Copenhagen interpretation), for Everett simply postulates a universal quantum
mechanical wave function. Moreover, Everett eliminates the baffling duality of
temporal development, one continuous and unitary governed by the Schr\"odinger
equation, and the other discontinuous and stochastic occurring when a reduction of
the wave packet occurs, since he denies the occurrence of the latter. The second is
that Bell recognizes some kinship between the Everett interpretation and the
hidden variables theory of de Broglie and Bohm, as no one previously seems to have
done, and the exploration of this kinship allows him to highlight certain virtues of
the latter. Everett's uncompromising adherence to quantum mechanics drives him
to a radical account of the measuring process:  If $\bigl\{\phi _n\bigr\}$ are a
complete set of eigenstates of an observable $A$ of an object, and the object
interacts with a measuring apparatus that is accurate in the sense of ``recording''
the eigenstate by entering a corresponding state $\chi _k$ whenever the object is
prepared in a definite
$\phi _k$, then the final state of object plus apparatus, when the object is prepared
in the superposition $\sum c_n\phi _n$, is   $\sum c_n\phi_n\chi_n$. For
Everett the final superposition is unequivocally interpreted as ``and'', not as ``or''.
He introduces the terminology ``relative state'' and says that relative to the
apparatus state $\chi_k$ the object state is  $\phi _k$, but there is no selection
from among those $\bigl\{\phi_n\bigr\}$ that occur with a nonzero coefficient in
the initial superposition. Each ``branch'' in the final state of object plus apparatus is
as real as any other. Does branching occur only when there is a measurement
process, the branches being labeled by the eigenstates of the ``pointer reading'' of
the apparatus, or does it occur relative to an arbitrary basis in the Hilbert space
associated with the apparatus? Bell finds an ambiguity in Everett's text on this
matter, and he raises objections to either possible answer. If the former, then
\begin{quote}
	Everett is indeed following an old convention of abstract quantum
measurement theory -- 	that the world does fall neatly into such
pieces -- instruments and systems\plips I think 	that fundamental
physical theory should be so formulated that such artificial 	divisions
are manifestly inessential. In my opinion Everett has not given such
a 	formulation, and de Broglie has.\quad\it\cite{n49} 
\end{quote}
If the latter answer is
accepted, with relative states defined for arbitrary bases, then ``it
becomes obscure to me that any physical interpretation has either
emerged from, or been imposed on, the mathematics.'' (Ref.~\cite{n26}, footnote 9).
Returning to the first answer regarding branching, we note that Bell
finds implicit in it a curious view of the past:
\begin{quote}
[I]n our interpretation of the Everett theory there is no association of
the particular 	present with any particular past. And the essential
claim is that this does not matter 	at all. For we have no access to the
past. We have only our `memories' and our 	`records'. But these
memories and records are in fact present  phenomena\plips 
	Everett's replacement of the past by memories is a radical solipsism
-- extending to 	the temporal dimension  the replacement of
everything outside my head by my 	impressions, of ordinary
solipsism or positivism. Solipsism cannot be refuted. But 	if such a
theory is taken seriously it would hardly be possible to take anything
else 	seriously. So much for the social implications.\quad\it\cite{n26}  
\end{quote}

\subsubsection*{(iii) The Hidden Variables Theory of de Broglie and
Bohm}

           In 1927 Louis de Broglie~\cite{n50} proposed a specific hidden
variables interpretation that was modified and amplified by David
Bohm~\cite{n30} in 1952. (We shall not analyze the similarities and
differences of the theories of the two men, but rather shall focus on
what we shall call ``BB'', which is Bell's own exposition of their
combined work.) BB applies quantum mechanics to the entire physical
world, and hence admits a universal wave function. Their hidden
variables are the positions of all particles, all positions being
objectively definite. The universal wave function evolves under the
Schr\"odinger equation, uninfluenced by the positions of particles. If
this wave function is expressed in the polar form
\numeq{e17}{
                                                  \psi (\vec q, t) =
R(\vec q, t)\exp [iS(\vec q, t)/\hbar] 
}
where $\vec q$ is position in
configuration space, $R$ and $S$ being real-valued functions, then BB
postulates that the generalized momentum conjugate to $\vec q$ is
\numeq{e18}{
   \vec p =\grad_q S\ .  
 }
 $\psi$ is often called ``the
guiding field'' and Eq.~\refeq{e18} ``the guiding equation''. Although
positions are the only explicit hidden variables, BB is able to ascribe,
with the following strategy, a definite result to a physically described
measurements of any quantity  represented in quantum mechanics
by a self-adjoint operator: The quantity of interest is coupled with an
apparatus whose ``pointer reading'' is assumed to be a position (hence
a component of $\vec q$), in such a way that an eigenstate of the quantity is
strictly correlated with a value of this position at the completion of the
measurement. Definiteness of $\vec q$, which is postulated by the theory,
entails a definite inference regarding the value of the measured
quantity.  The statistical predictions of BB are the same as those of
standard quantum mechanics if one supposes that the Born rule for
probability density,
\numeq{e19}{
                                              \prob(\vec q,t _0) =
|\psi(\vec q,t_0)|^2 
}
holds for some initial time $t_0$, its
holding for all other times being a consequence of the dynamics.

    Bell makes a number of complimentary remarks about BB. The
theory does not take an anthropocentric concept like observables as
fundamental, but postulates an ontology consisting of  (i) a universal
wave function, thought of as physically existing like the
electromagnetic field, and (ii) the positions $\vec q$, which are the ``beables''.
Indeed, these beables ought not be thought of as $\it hidden$  variables,
because positions are manifest quantities, whereas the wave function
is shadowy and shows its existence only via its effect upon the
positions. There is only one dynamical law governing the wave
function and no mysterious ``reduction of the wave packet'' to account
for a result of measurement. There are definite measurement results,
but they are the values of certain components of $\vec q(t)$, whose temporal
development depends only on an initial value, the wave function, and the
guiding equation. Although BB shares with Everett's interpretation the
virtue of avoiding two different laws for the evolution of the wave
function, it differs -- to its advantage -- by attributing a definite
configuration to the world instead of a multiplicity of branches with
different configurations. It is also a virtue of BB that nonlocality enters
in a natural way:  When the wave function of the world -- or of a small
number of systems in an idealizing approximation -- is an entangled
state, the components of the generalized momentum $\vec p$ in the guiding
Eq. \refeq{e18} can be correlated even though the corresponding
components of $\vec q$ are far separated:
\begin{quote}
	That the guiding wave, in the general case, propagates not in
ordinary three-space 	but in a multidimensional-configuration space is
the origin of the notorious 	`nonlocality' of quantum mechanics. It is
the merit of the de Broglie-Bohm version 	to bring this out so
explicitly that it cannot be ignored.\quad\it\cite{n51}, last paragraph 
\end{quote}

Finally, the essential idea of BB is more general than the detailed
exposition by its authors, for there is no reason in principle to restrict
the beables to particle positions,  suspiciously reminiscent of classical
mechanics. There are other possible choices of beables, suggested by
recent discoveries in elementary particle physics, which would be
coupled with the universal wave function by an appropriate guiding
equation. In ``Beables for Quantum Field Theory''~\cite{n44} -- dedicated to
Bohm -- Bell suggests a version of BB that takes fermion number
density as the beables, since that ``includes the positions of
instruments, instrument pointers, ink on paper\lips  and much much
more.'' The resulting theory is BQFT (``de Broglie-Bohm beable
quantum field theory''), which has to be compared with OQFT 
(``\ts`ordinary' `orthodox' `observable' quantum field theory''). There is
good agreement between BQFT and two versions of OQFT that Bell
presents.  In spite of this empirical parity, at least FAPP, Bell prefers
BQFT conceptually, because it is designed to provide a detailed
temporal account of the measurement process.

After thus making his case for a suitably modified BB, Bell concludes
with a strong reservation.
\begin{quote}
	 BQFT agrees with OFQT on the result of the Michelson-Morley
experiment, and 	so on. But the formulation of BQFT relies heavily on
a particular division of space-time into space and time. Could this be
avoided?\quad\it\cite{n44}, Sect.~5 
\end{quote}
 After a brief discussion of ways to achieve Lorentz invariance, Bell
concludes skeptically:
\begin{quote} 
I am unable to prove, or even formulate clearly, the
proposition that a sharp 	formulation of quantum field theory, such as the
one set out here, must disrespect 	serious Lorentz invariance. But it seems to
me that this is probably so.\quad [ibid.]
\end{quote}    

\subsubsection*{(iv)  Theories of Spontaneous Reduction of the Wave
Packet}
    
Bell was aware of a number of proposals to solve the measurement
problem by modifying the time-dependent Schr\"odinger equation,  by
the addition of small nonlinear terms or by introducing stochasticity
into the dynamics (see footnote 6 of Ref.~\cite{n26}). The detailed proposal of this
kind that impressed him most was that of G.C. Ghirardi, A. Rimini, and
T. Weber (GRW)~\cite{n52}, which he discusses in detail in ``Are There
Quantum Jumps?''~\cite{n53}. GRW postulate no hidden variables but
consider the quantum state to be a complete description of a physical
system. The quantum state satisfies Bell's criterion for a ``beable''
formulation of quantum mechanics, because they describe the
measurement process explicitly and clearly in terms of the
fundamental dynamical law;  their paper is, in fact, entitled ``Unified
Dynamics of Microscopic and Macroscopic Systems.'' GRW's innovation
is a stochastic reduction which may occur for any particle at random
infrequent times (the mean waiting time for reduction to occur being a
new constant of nature $\tau$, clearly large and postulated by GRW to be
$10^{15}$~s).The stochastic reduction is described in the position
representation, and it can occur even when the particle is part of a
composite system described by an entangled state,
\numeq{e20}{
 \psi( \vec r _n, \vec x',t)  =  \sum_i c _{ni} (t)\phi _i(\vec r _n)\Phi_i(\vec x')
}
where $\vec r_n$ is the position of the $n^{\mathrm{th}}$ particle, $\vec x'$ is the
configuration of all particles except the $n^{\mathrm{th}}$, and $i$ is an index in a
superposition of terms. When the stochastic reduction happens for the
$n^{\mathrm{th}}$ particle, the result is to replace every $\phi _i(\vec r_ n)$  by 
$j(\vec r - \vec r_n)\phi _i(\vec r _n)/R_n(\vec r)$, where $j$ is a ``jump factor'',
Gaussian for convenience with width~$a$ of the order of $10^{-5}$~cm, and $R_n$ is
a normalization factor. After a jump the $n^{\mathrm{th}}$ particle will be well
localized, since all wave functions
$\phi _i$ with supports not close to $\vec r$ will be annihilated by the jump factor.
One virtue of this scheme is that for a single particle or a system of few particles,
the stochastic modification of the Schr\"odinger equation causes such infrequent
jumps that they will not be experimentally noticeable. On the other hand, for
a system with a large number of particles, of the order of $10^{20}$, the
mean waiting time for reduction of some particle or another of the
system will be small, of the order of $10^{-5}$~s, and therefore -- because of
the algebra of entanglement -- the position of the center of mass will
quickly be localized within a region of diameter $10^{-5}$~cm. Thus a
macroscopic system will not have a grossly unlocalized position more
than a short time, and the pointer reading is quickly definite. Classical
dynamics is approximately recovered for macroscopic systems by an
explicit calculation based upon the dynamics of microscopic systems.
Although GRW's experimental disagreements with standard quantum
are small, they are nonzero and in principle testable (but so far no
feasible test has been devised).

The most troublesome feature of GRW, according to Bell, is the
apparent action-at-a-distance that occurs when one of the particles of
a composite system in an entangled state undergoes a spontaneous
jump. A particle remote from the one undergoing the jump may be
constrained to be localized, without any locally describable causal
process. To be sure, no message can be transmitted by this correlation,
but Bell's reasons for thinking  this fact is not crucial were
presented in Section~\ref{ss:E} above. What concerns him is whether the theory can
be modified so as to be Lorentz invariant in spite of spooky correlations.
The question cannot be directly investigated in GRW's theory, which is
nonrelativistic, but Bell does answer a related question. He proves
that if a different time coordinate is used for each particle, a relative
time translation invariance holds in this theory. He therefore
concludes tentatively, but somewhat optimistically, concerning the
prospects of further developments of GRW:
\begin{quote} 
I see the GRW model as a very nice illustration of how quantum
mechanics, to 	become rational, requires only a change which is small
(on some measures!). And I am particularly struck by the fact that
the model is as Lorentz invariant as it could be in the nonrelativistic
version. It takes away the ground of my fear that any exact 
formulation of quantum mechanics must conflict with fundamental
Lorentz 	invariance.\quad\it\cite{n53}, Sect.~5 
\end{quote}
 It is regrettable that Bell did
not live to develop further a program that he found so promising.

\section{Field Theory and Elementary Particle Physics}\label{s3}

John Bell's public profession(also his employment) from the mid-1950s was
that of a theoretical nuclear and later particle physicist, a progression
that reflects the historical development of the subject.  The framework
within which he worked was quantum field theory.  During his time,
two competing ideas were also popular among particle theorists: 
S-matrix/Regge theory in the early years, string theory in later years.
Apparently these did not appeal to Bell, who remained within field
theory for all his investigations.  His first paper in this area, in 1955, 
followed  five active years of research  in accelerator physics, which
produced 21 publications. Drawn from his doctoral thesis at
Birmingham University, it addresses an important and classic issue.
Bell entitled the work ``Time reversal in field theory"~\Cite{n1}.  In
fact he established the PTC theorem, which states that any local,
relativistically invariant quantum field theory is invariant against the
simultaneous inversion of space-time coordinates and reversal of all
charges.  Unknown to him this had already been shown the year
before by L\"uders using an argument markedly different from
Bell's~\Cite{r2}.  Yet  Bell is  hardly ever credited for his independent
derivation, presumably because he was not in the circle of formal field
theorists (Pauli, Wigner, Schwinger, Jost, etc.)\ who appropriated and
dominated this topic.  Today Bell's ``elementary derivation" is more
accessible than the formal field theoretic arguments. The subject of
time reversal remained an important theme in his subsequent
research~\Cite{r3}, especially when it became clear that time inversion
(T) (unaccompanied by space inversion and charge conjugation) is not
a symmetry of Nature, and neither is space inversion conjoined with 
charge conjugation (PC) (unaccompanied by time inversion). Together
with his friend, the experimentalist Steinberger, he wrote an influential
review on the phenomenology of PC-violating experiments~\Cite{r4},
and with Perring proposed a ``simple model" theory to explain that
effect~\Cite{r5}.  They postulated the existence in the universe of a
hitherto unobserved long-range field, which would catalyze the PC
symmetry breaking.  Though speculative, the suggestion was truly
physical, hence falsifiable, and it was soon ruled out by further
experiments. Nevertheless, it remains a bold and beautiful idea that
continues to intrigue theorists.

Bell's immediate postdoctoral research concerned nuclear physics, and
he worked closely with Skyrme on nuclear magnetic
moments~\Cite{r6}, though apparently not on Skyrme's prescient ideas
on solitonic models for the nucleons.  Bell contributed variously to the
many-body theory of nuclei~\Cite{r7}, notably the derivation with
Squires of an effective one-body ``optical" potential for the scattering
off a complex target~\Cite{r8}. The nuclear physics work, though not
producing any breakthrough in that field, prepared him well for
dealing with problems that he would eventually encounter at CERN,
involving nuclear and particle processes.  Examples of this later work
are his nuclear optical model (again!)\ for pi-mesons ~\Cite{r9}, and
investigations of neutrino reactions with nuclei~\Cite{r10}, where the
hypothesis of a ``partially conserved axial vector current" (PCAC)
connects neutrino processes with pi-meson emission.

In 1960, Bell joined CERN and remained there to the end of his life. 
Although primarily a center for particle physics accelerator
experiments, CERN houses Europe's preeminent particle theory group,
and Bell became active in that field. Typically particle theorists are
divided into phenomenologists -- people who pay close attention to
experimental results, and interact professionally with experimentalists
-- and formalists, who explore the mathematical and other properties
of theoretical models, propose new ideas for model building, and
usually are somewhat removed from the reality of the experiments.  
Although Bell's time-reversal paper~\Cite{n1} belongs forcefully in the
formalist category, at CERN he was very much also a particle physics
phenomenologist, drawing on his previous experience with nuclear
physics. Indeed, with characteristic conscientiousness,  Bell felt an
obligation to work on subjects related to the activities of the
laboratory. But his readiness to discuss and study any topic in physics
ensured that he would pursue highly theoretical and speculative
issues as well.

Theoretical understanding of particle physics at that time was
hampered by two obstacles.  No single model had been identified as
giving a correct account for the fundamental interactions of
elementary particles, with the exception of electromagnetic
interactions, which were adequately described by quantum
electrodynamics. Moreover, competing models could not be assessed
because their extremely complicated dynamical equations could not be
solved, so the predictions of the models were unknown and could not be
compared with experiments.

To overcome this impasse, Gell-Mann advanced the idea of ``current
algebra'', with which one could obtain explicit and testable results. 
Current algebra is a particle physics/quantum field theory reprise of
an old technique used in quantum physics for atoms, namely, the
Thomas-Reiche-Kuhn sum rule, or the Bethe energy loss sum
rule~\Cite{r11}.   Here one identifies relevant operators whose matrix
elements govern transitions between atomic levels, and the total
transition probability is a sum over all final-state levels.  In favorable
circumstances, the operators are manipulated by using the operator
quantum commutation relations, which are deduced from the
fundamental canonical commutators, and the sum can be evaluated
using the completeness of states.  All this can be achieved without
determining individual wave functions, which would involve the
daunting task of solving completely the interacting Schr\"odinger
equation; indeed, even explicit knowledge of the potential function
governing the interactions is not needed.

In the particle physics analogue, the relevant operators governing
transitions are the relativistic 4-vector and axial 4-vector currents,
generalizations of the electromagnetic 4-vector current, but carrying 
internal symmetry group $[SU(2)$ or $SU(3)]$ labels.  In an article in the
same volume of the now-defunct journal $\it{Physics}$ in which Bell
published his famous EPR paper, Gell-Mann postulated a form for the
commutators of current components using a hypothesis about
canonical structure and plausible ideas from group theory~\cite{r12}.  Also it was
necessary to know the covariant divergences of the 4-vector and axial
4-vector currents.  One took the vector current divergence to vanish,
that is, that current is conserved just as is the electromagnetic
4-vector current, while the axial 4-vector current was taken to be
partially conserved  (PCAC) and related to pi-meson processes.

Gell-Mann's proposal~\Cite{r12} successfully bypassed the two
obstacles:  ignorance about specific dynamics, and inability to unravel
proposed  dynamical equations.  Testable results were soon obtained
by Adler, Fubini, Furlan,  Weisberger, and others~\Cite{r13},  and the
predictions of current algebra theory appeared to agree very well with
experiment.  Consequently almost all particle physics theorists began
researching the foundations, extensions, and applications of
Gell-Mann's current algebra proposal.  So also did Bell, and it is within
this area that he made his principal contributions to particle physics.

While the canonical formalism in a quantum field theory provides one
starting point for deriving the current algebra, it suffers from the
shortcoming that canonical commutation relations are postulated for
the unrenormalized operators. But quantum field theory is notoriously
polluted by various infinities, which must be renormalized, and it is
not clear whether relations obtained by manipulating unrenormalized
quantities accurately describe what the model entails.  Bell illuminated
these issues by studying the commutators in a ``trivial" completely
solvable but unrealistic model, which nevertheless requires
renormalization (Lee model)~\Cite{r14}. He demonstrated that indeed
canonical relations must be used with caution, because in various
circumstances they will lead to values for particle physics sum rules
(analogs of the atomic physics Thomas-Reiche-Kuhn or Bethe sum
rules) that do not agree with the explicit summation of the explicitly
calculated amplitudes. Also, under the influence of related work by his
friend and colleague Veltman~\Cite{r15}, Bell proposed a different
basis for the current algebra.   Rather than relying on canonical
commutation relations for unrenormalized operators, he showed that
invoking the non-Abelian gauge principle, which lies at the heart of
Yang-Mills theory -- a subject Veltman was intensely studying -- will
also yield the desired current commutation relations~\Cite{r16}.

After the successes of current algebra built on the $SU(2)$ or $SU(3)$
group, attempts were made to employ other groups $[U(4), U(6), SU(6),
SL(6,C), U(6)\times U(6), U(12)$, etc.].  The extended theory had an
initial success in predicting the ratio between neutron and proton
magnetic moments to be $-2/3$, which agrees well with experiment. 
But soon it became evident that a consistent relativistic theory could
not incorporate such groups as symmetry groups, and Bell contributed
to the critique~\Cite{r17}.

When this generalization fell out of favor, most people left the subject,
but Bell kept an eye on the topic, owing to his longtime interest (since
his time with Skyrme) in nuclear magnetic moments, and his belief
that the good experimental agreement should not be accidental.  Thus
when the subject reappeared years later in the guise of the ``Melosh
transformation", Bell researched and lectured on this new
direction~\Cite{r18}, but results remained inconclusive.

In addition to formal, theoretical investigations and critiques of
current algebra, Bell derived and refined various useful relations
relevant to experiment:  beta-decay of pi-mesons~\Cite{r19},
low-energy Compton scattering~\Cite{r20}, soft pi-meson
emission~\Cite{r21}. But in these studies he also encountered results
that failed to agree with experiment, even though the derivations
made use of what appeared to be the most reliable aspects of the
theory.   First there was the current algebra and PCAC analysis that
seemed to forbid the decay of the
eta-meson into three pi-mesons, even though the process is seen
experimentally~\Cite{r22}.  This complemented the result of
Sutherland and Veltman~\Cite{r23} that a very direct application of
current algebra and PCAC prohibits the decay of the neutral pi-meson
into two photons, again contradicting experimental observation. 
Although these were small defects of the generally successful current
algebra story, John Bell did not put them aside, since he would not
abide imprecision and incompleteness in physical theory.

The two streams of Bell's current algebra research -- investigations on
the reliability of the postulated algebraic structures, and attention to
apparent failures of the theory -- came together on what turned out to
be his most far-reaching contribution to particle physics, published in
his most-quoted scientific paper.  Responding to a request by Jackiw
(who in 1967--8 was a visiting scientist at CERN) for a research
problem, Bell suggested analyzing current algebra's failure in the
pi-meson/two-photon  process.  This was a vexing puzzle, because the
theoretical argument prohibiting the decay seemed to be direct and
elementary.  No new ideas were forthcoming, until a casual
observation (over coffee in the CERN caf\'e) by Steinberger, Bell's
experimentalist collaborator on their PC review article~\Cite{r4},
pointed the way.   Steinberger remarked that years earlier he had 
computed the amplitude in a once popular model field theory, and
gotten a nonvanishing result, which moreover agreed well with
experiment~\Cite{r24}.  Bell and Jackiw realized that Steinberger's
calculation would coincide with what one would find in the
$\sigma$-model, a field theory constructed by Gell-Mann and L\'evy
to exhibit explicitly current algebra and PCAC~\Cite{r25}. So here was
the test:  On the one hand, a direct calculation of the decay amplitude
in the $\sigma$-model should reproduce Steinberger's nonvanishing
result.  On the other hand, an indirect calculation based on current
algebra/PCAC, which appear to be present in the $\sigma$-model,
should give a vanishing result.  The resolution of this conflict would
display what is happening, and is reported in their paper, ``A PCAC
Puzzle: $\pi^0 \to \gamma\gamma$ in the $\sigma$-Model"~\Cite{r26}.

The direct calculation involves the now famous fermion triangle graph
$$
\BoxedEPSF{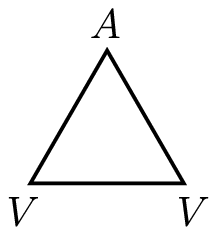}
$$ where the bottom two vertices denote electromagnetic vector
currents to which the two photons couple, while in the apex there is
the axial vector current, which according to PCAC governs the
pi-meson coupling.  The lines joining the vertices denote fermion
propagators, which in the Steinberger calculation were
protons~\Cite{r24}, while in contemporary theory they are quarks.

The three-current amplitude is a three-index tensor, carrying a
space-time index for each of the three currents.  Direct evaluation of
the Feynman graph reproduces Steinberger's nonvanishing decay
amplitude~\Cite{r24}.  On the other hand, current algebra and PCAC
demand that with massless fermions propagating between the
vertices, the three-index tensor should be transverse in each of its
indices.  These transversality conditions are examples of
Ward-Takahashi identities satisfied by field-theoretic amplitudes.
Generically, the task of current algebra and PCAC was to establish such
identities, from which physically relevant consequences could be
drawn.  In the pi-meson/two photon application, the threefold
transversality of the three-current amplitude  implies that the
pi-meson decay amplitude vanishes~\Cite{r23}.  Since the explicit
calculation provides a nonvanishing amplitude, the Ward-Takahashi
identity must fail and indeed the explicit calculation of the triangle
produces a nontransverse result~\Cite{r26}.

This observation put into evidence a previously unsuspected quantum
field theoretic phenomenon:  With massless fermions, the axial vector
current appears to be conserved.  (That is why the three-current
amplitude is expected to be transverse in the axial vector index, just
as it is in the vector indices, owing to the conservation of the vector
electromagnetic current.)  Current conservation is evidence that a
symmetry holds: conservation of the electromagnetic current is
correlated with gauge invariance.  With massless fermions, a
conserved axial vector current would indicate that a further symmetry
is present: the so-called chiral symmetry.  Chiral symmetry is
understood in the following fashion.  Massless spin-$\frac {1}{2}$ fermions can
be viewed as consisting of two species,  which do not mix. In one
species the spin is aligned along the direction of propagation; in the
other, the spin points opposite to the propagation. Chiral symmetry
then is the ability of performing separate and independent gauge
transformations on each species.  Noninteracting fermions, described
by a massless Dirac equation, do indeed exhibit this chiral symmetry. 
Superficially, it appears that interaction with photons should preserve
the chiral symmetry.  However, the triangle graph calculation
demonstrates that this is not so~\Cite{r26}.  The technical reason for
this unexpected violation of the symmetry is that the integral that
evaluates the triangle graph is singular (divergent) and a prescription
must be given how to handle the singularity to get a noninfinite
result.  It turns out that no matter what prescription is given, chiral
symmetry is lost.  Therefore, if we accept that quantum field theoretic
infinities are a necessary consequence of the quantization procedure,
the Bell and Jackiw discovery established a novel method of symmetry
breaking, called anomalous or quantal symmetry breaking, which
arises from the quantization procedure itself: symmetries of an
unquantized theory need not survive quantization. This new mode of
symmetry breaking  supplements previously known possibilities:
symmetry breaking by nonsymmetric dynamics or symmetry
breaking by energetic instabilities, as in spontaneous symmetry
breaking with the Higgs mechanism. Moreover, this quantal symmetry
breaking mechanism also violates the correspondence principle of
quantum physics~\Cite{r27a}.

The situation can be summarized by the anomalous divergence
equation. Rather than being conserved, the axial vector current has a
nonzero divergence, given by
\numeq{e21}{
\partial_\mu J^\mu_A \propto {}^*\!F^{\mu\nu} F_{\mu\nu}\ .
}
   On the left there is the divergence of the axial vector current
$J^\mu_A$, on the right is the anomaly responsible for the quantum
mechanical nonconservation. Here $F_{\mu\nu}$ is the electromagnetic
Maxwell tensor, 
${}^*\!F^{\mu\nu}$ its dual equal to $\frac12
\epsilon^{\mu\nu\alpha\beta} F_{\alpha\beta}$, and their product
coincides with the scalar product of the electric and magnetic fields
with which the fermions interact.  The proportionality constant in the
above anomalous divergence equation is determined by the number
and charges of the fermions and fixes the strength of the
pi-meson/two-photon amplitude.

The implications of this symmetry breaking phenomenon are many,
and they are widely spread in physical theory.  Here is a list.
\begin{enumerate}

	\item  Working independently, and slightly later, Adler found that
the axial vector current in quantum electrodynamics behaves
similarly. When he learned of the Bell-Jackiw considerations about the
pi-meson/two-photon amplitude, he complemented their work, and
proved (with Bardeen) that no other graphs contribute to the
anomaly~\Cite{r27}. Therefore,  a precise value for the
pi-meson/two-photon amplitude can be predicted.  Numerical
agreement with the experimental value requires that if the fermions
propagating in the legs of the triangle graph are quarks, carrying conventional
quark charges, there must be three species -- ``colors".

	\item  Current commutators reflect the underlying symmetry of the
theory.  If the symmetry is broken quantum mechanically, there must
be quantal correction to current algebra.  Such correction were found
and they provide an alternative viewpoint why Ward-Takahashi 
identities fail:  anomalous divergences of currents and anomalous
commutators are two sides of the same coin~\Cite{r28}.

	\item  It soon became apparent that anomalous divergences can arise
in other currents as well.  However, if gauge fields couple to currents,
consistency of the gauge principle requires that such gauge currents
be anomaly free.  This can be arranged, provided the number and
charges of participating fermions are such that their contributions to the
proportionality constant in the anomalous divergence of the gauge
current sum to zero.  In the present day ``Standard Model" of
electroweak interactions this cancellation of anomalies requires equal
numbers of quarks and leptons, with charges taking precisely the
values of the Standard Model~\Cite{r29}.  Moreover, anomalies also
endanger consistency of string theories, and the demand of anomaly
cancellation strongly limits the possible string theories.  This limitation
led to a revival of string theory research~\Cite{r30}.

	\item  Polyakov and his collaborators realized that the quantity
${}^*\!F^{\mu\nu}F_{\mu\nu}$ and its non-Abelian generalization are
topological entities, whose four dimensional integral measures the
topological twist in the underlying gauge fields.  This is the
Chern-Pontryagin number. They identified fields with nonvanishing
twist -- the instantons -- even though they made no reference to the
anomaly, which evidently was largely unappreciated at that time in
the Soviet Union~\Cite{r31}.

	\item  't~Hooft, a student of Bell's friend Veltman, and well instructed
by him about Bell's anomaly work, showed that in the Standard Model
certain combinations of fermion number currents do not couple to
gauge fields and are afflicted by the anomaly. As a consequence
protons can decay, but with an exponentially vanishing small
probability, thus not endangering the stability of our
world~\Cite{r32}.  Related to this was the observation that owing to
the anomaly, there is the possibility of quantum mechanical tunneling
in the Standard Model field theory.  This creates a Bloch-like  band
structure, where states are labeled by a hitherto unsuspected
PC-violating parameter -- the
$\theta$~vacuum angle~\Cite{r33}.  How to fix the magnitude of this
$\theta$-angle is still an open question.

	\item  Discussion of anomalies brought physicists into contact with
mathematical entities, like the previously mentioned
Chern-Pontryagin term, and the Chern-Simons term~\Cite{r34} (whose
exterior derivative gives the Chern-Pontryagin entity). As it happened,
mathematicians were studying precisely these same topics, at roughly
the same time.  Thus, the anomaly seeded a remarkable
physics/mathematics collaboration and cross-fertil\-i\-zation, which is still
flourishing, especially in string theory research~\Cite{r35}.

	\item  It was appreciated that another symmetry, which relies on
masslessness, is beset by anomalies.  This is scale and conformal
invariance, which would require the energy momentum tensor to be
traceless.  But in fact superficially scale and conformally invariant
theories (like the Standard Model in the absence of its Higgs sector)
acquire upon quantization a nonvanishing trace -- a trace anomaly --
thereby breaking quantum mechanically the scale and conformal
symmetries~\Cite{r36}.  This of course is very fortunate, because
Nature certainly is not scale invariant, and could not be described by a
scale-invariant theory. Moreover, it was appreciated that the proper
way to deal with anomalously broken scale and conformal symmetry
in quantum theory is through the Gell-Mann--Low renormalization
group~\Cite{r37}.

	\item  Beyond particle physics theory, in condensed matter theory
and in gravity theory, physicists have realized  that a mathematical
discussion of physical effects can have a topological component, which
is related to  structures first seen in quantal anomalies.  Examples are
descriptions of fractional fermions by spectral flow of the Dirac
operator, the theory of the quantum Hall effect based on the
Chern-Simons term.

\end{enumerate}

The direct physical relevance of the axial vector anomaly in accounting
for the decay of the neutral pi-meson, and in explaining quark and
lepton patterns that ensure absence of anomalies in gauge currents, is
evidence that not only theoretical/mathematical physicists but also
Nature knows and makes use of the anomaly mechanism.  Moreover,
the unexpected connections that the anomaly makes within physics
and with mathematics suggests that we are dealing with an as yet not
completely understood wrinkle in the mathematical description of
physical phenomena.

Once it was appreciated that the anomaly phenomenon is not merely
an obscure pathology of quantum field theory, many people wrote
many papers providing various and alternative derivations of the
result.  But not John Bell.  It seems that he was satisfied with what he
stated in his only paper on this topic~\Cite{r26}. He did follow the
subsequent developments and elaborations, but apparently he
preferred the ``simple" triangle-diagram calculation that started the
subject.  One reason for this is that he felt that the diagrammatic
approach, leading to a singular integral, very explicitly exhibits the
arbitrariness and ambiguity in the calculation, which can only be
resolved when some additional, external information is added.  For
example, the three-current triangle graph cannot be transverse in all
three vertices, but there is no information in which vertex
transversality fails since different methods for handling the
singularity produce different results.  A unique value, fixing the lack
of transversality in the axial vector vertex, emerges only after
requiring the vector vertices be transverse, because gauge particles --
photons -- couple to these vertices.  On the other hand in the standard
electroweak model, with the same triangle graph, the vector vertex
describes a nongauged fermion number current, and in this context
the anomaly can reside there, giving rise to  't~Hooft's proton decay
scenario~\Cite{r32}.

After the anomaly paper, Bell returned to more immediately practical
investigations like the already-mentioned neutrino-nucleus work with
Lewellyn-Smith~\Cite{r10}, and  the Melosh reprise of ``higher"
symmetries~\Cite{r18}.

This was also the period when non-Abelian gauge theories became
accepted as the correct field theoretic description of fundamental
processes in the Standard Model, with the Higgs mechanism providing
a mass  for the carriers of the weak interaction forces. Although he
was an early proponent of non-Abelian gauge symmetry in current
algebra~\Cite{r16}, Bell did not work extensively in this newly
expanding field, writing only on two topics.  He showed that the Higgs
mechanism ensured good high-energy behavior in theories with
massive vector fields (such as those that carry the weak interaction
forces)~\Cite{r38}.  Then with Bertlmann and others he examined
various ideas about bound states in quantum chromodynamics -- the
hadronic sector of the Standard Model~\Cite{r39}.  These were the last
particle physics studies carried out by Bell. The final scientific papers
that he wrote all deal with accelerator physics and with his quantum physics
``hobby", which late in his life became widely appreciated.  There were many
demands on him to explain and develop these ideas.

Two other investigations are noteworthy.  As if anticipating the great
excitement felt these days about experimental/theoretical
discrepancies in the anomalous magnetic moment of the muon $a_\mu
=
\frac12(g_\mu-2)$,  Bell and de~Rafael obtained an upper bound on
the hadronic contribution:
$a_\mu\bigr|_{\mathrm{hadronic}}$ 
$<10^{-6}$~\Cite{r40}. This is certainly satisfied by today's accepted
value,
$a_\mu\bigr|_{\mathrm{hadronic}}$ $\sim 7\times10^{-8}$.  But a
change in the accepted number consistent with the Bell-de~Rafael
bound could very well affect the experimental/theoretical discrepancy
$\Delta\, a_\mu \sim 4\times10^{-9}$.

A peculiar effect seen in quantum field theory, which challenges
conventional ideas, and has a realization in condensed matter physics
attracted Bell's interest.  It was observed that the measured
value for the number operator of a fermion moving in the background
of a topological soliton, as for example created by a domain wall in a
solid-state substance, would be a fraction~\Cite{r41}. Moreover, it was
alleged that this phenomenon is physically realized in polyacetylene --
a one-dimensional polymer~\Cite{r42}.  One naturally wonders
whether the observed fraction is an expectation value, or an
eigenvalue without fluctuations.  Only in the latter case would this
represent a truly novel and unexpected phenomenon.

On a visit to India, Bell's host Rajaraman described to him this effect,
but Bell doubted that the fraction could be a sharp observable. 
Nevertheless, he wanted to find out and in two papers they
established that indeed the fractions were eigenvalues of a number
operator, ``which is defined with some sophistication"~\Cite{r43}.

This last-mentioned work illustrates well John Bell's attitude to his
research on fundamental physical questions.  Rather than advancing
new theoretical models, his publications are infused with a desire to
know and explain existing structures, preferably in ``simple terms", in
a ``simple model" -- phrases that occur frequently in his papers. If
physicists come in two types, those who try to read the book of Nature
and those who try to write it, Bell belonged to the first category. He
was conservative when it came to speculative and unconventional
suggestions; he would prefer that unexpected contradictions not arise,
that ideas flow along clearly delineated channels.  But this would not
prevent him from establishing what exactly is the case and accepting,
albeit reluctantly, even puzzling results.  This tension is seen in the
anomaly paper, where after describing the phenomenon in the first
part, a later section is devoted to an attempt at removing the anomaly,
saving the chiral symmetry~\Cite{r26}. (Years later it was understood
that this attempt yields a construction with scalar fields of the
Wess-Zumino anomalous effective Lagrangian, which can be used to
compensate for the fermion-induced anomaly~\Cite{r44}.)  Even in his
quantum mechanical investigations, Bell would have preferred to side
with the rational and clearly spoken Einstein rather than with the
murky pronouncements of Bohr.  But once he convinced himself where
the truth lies, he would not allow his investigations be affected by his
inclinations, even if he remained disturbed by their outcome. Such a
commitment to ``truth" -- as he saw it -- marked John Bell's activity in
science and in life.

\section{Accelerators}\label{s4}

From the time that Bell came to Harwell in 1949 until his leave for graduate work
at the University of Birmingham in 1953--4 his research almost entirely concerned
accelerator design, in a group directed by William Walkinshaw.  The first
twenty-one items in his bibliography, from 1950 to 1954, concerned aspects of
accelerator design.  Of these, only two were published; the others are internal
A.E.R.E. reports at Harwell, but these reports were considered authoritative and
were much consulted and cited. He continued to  make contributions to accelerator
design during his entire career.  Since neither of the authors  of this essay is an
expert on accelerators, we shall do little more than note the topics of Bell's
research.  But this part of his work must be kept in mind even by readers primarily
interested in his contributions to foundations of quantum mechanics and high
energy physics, because it shows the practicality and concreteness of his mind,
which undoubtedly deeply influenced even his highly theoretical and philosophical
thinking.

In the Royal Society memorial essay on Bell by P.G. Burke and I.C.
Percival~\cite{n98}, Walkinshaw is quoted concerning collaboration with him:
\begin{quote} 
We in Malvern had finished our work on small electron machines, so
had turned our attention  to high-energy machines.  The situation was
fluid and we looked at all sorts of possibilities.  One of these was a disc-loaded
waveguide for electron accelerators. This was the first project that John worked on
with me~\cite{n99}\lips  another possibility was a high-energy proton linac, and we
looked at variety of waveguide structures\ldots Here was a young man of high
calibre who soon showed his independence on choice of project, with a special
liking for particle dynamics.  His mathematical talent was superb and elegant.
\end{quote} 
Mary Bell~\cite{n100} pointed out that his mathematical skill enabled
him to carry design problems to the point where the calculations required could be
performed on the primitive desk calculators available at that time.

When the group began to consider the design of a proton synchrotron, Bell  studied the application of the
strong focusing principle, which was the topic of his first published paper~\cite{n101}.  According to the
memorial essay,~\cite{n98} he ``wrote a seminal report on the algebra of strong focusing~\cite{n102}\ldots read by all accelerator designers of the day."  Other topics dealt with in internal
reports were ``Linear Accelerator with Spiral Orbits", the ``Stability of Perturbed
Orbits in the Synchrotron," and ``Linear Accelerator Phase Oscillations."

The collaboration between Mary and John Bell on accelerator problems, that had
commenced at Harwell, continued at CERN although he was in the Theory Division
and she was in the Accelerator Research Group.  They wrote several papers on
electron cooling~\cite{n103,n104} when CERN's Super Proton Synchrotron was being
designed, and even though a less expensive stochastic cooling method was adopted
for that accelerator, the Bells' cooling method (also proposed by some Soviet
physicists) was later used for the LEAR ring. They also published several papers
together~\cite{n105,n106} on Brems\-strahlung, a practically important
phenomenon because it contributes substantially  to energy loss in
electron-positron linear colliders.  Richard Hughes, a visitor at CERN who
collaborated with Bell, commented~\cite{n107}, ``It was perhaps from his work in
this field that John acquired his rigorous understanding of classical mechanics and
special relativity" (though we suspect an earlier acquisition of this expertise), and
he cited published lectures by John Bell on Hamiltonian mechanics at the CERN
Accelerator School~\cite{n108}.

One of the most remarkable of Bell's achievements was the result of combining
considerations from accelerator physics with quantum field theory.  In 1976
William Unruh~\cite{n109} published the derivation of a beautiful ``effect" (which
became known as the ``Unruh Effect"):  An idealized detector uniformly
accelerated through the electromagnetic vacuum will experience radiation with the
spectral distribution of black body radiation at a temperature proportional to the
acceleration; specifically
\numeq{e22}{
kT_U= \hbar a/c\ .
}
 Jon Leinaas, who was a fellow at CERN in the early eighties, discussed this effect
with Bell, and asked whether elementary particles could be used as detectors of
this effective black body radiation~\cite{n110}.  They showed that existing linear
accelerators, which approximated the conditions of the Unruh effect, produce
effects too small to be seen experimentally.  However,
\begin{quote}
As suggested by John Bell, one might ask whether the temperature effect could be related to a polarization
effect which was known to exist for electrons circulating in a magnetic field.  It had already been established
that electrons in a storage ring polarized spontaneously, but not
completely\lips; the maximum polarization had been found to be
$P=0.92$.\quad\it\cite{n110}   
\end{quote}
They investigated this question and found a qualified positive answer.  Circulating electrons experience spin
transitions due to quantum fluctuations of the vacuum field,  just as transitions are induced in linearly
accelerated detectors.  But there are complications in the case of circular motion. The Thomas precession has
to be taken into account.  The excitation spectrum is not universal as in the linear Unruh effect but depends
upon characteristics of the detector.  There is also a narrow resonance involving vertical fluctuations and
the spin motion, with a peculiar influence on polarization:  as a function of $\gamma=(1-\beta^2)
^{- 1/2}$  the calculated polarization close to the point of resonance first decreases strongly,  then
increases to $0.99$ before decreasing again approximately to the observed saturation polarization~\cite{n111}.  These fine points enhance their extraordinary achievement of converting a highly theoretical
Gedanken experiment into a real experiment.

\newpage

\def\Journal#1#2#3#4{\emph{#1} {\bf #2}, #3 (#4)}
\def\add#1#2#3{{\bf #1}, #2 (#3)}
\def\Book#1#2#3#4{{\em #1}  (#2, #3 #4)}
\def\Bookeds#1#2#3#4#5{{\em #1}, #2  (#3, #4 #5)}
\def\Proc#1#2#3{\emph{#1},  #2 (#3)}

\let\J\Journal
\let\B\Book

\def\T#1{``#1'',\ }
\def\Ti#1{``#1'' in }

\def\NPB{Nucl. Phys.} 
\def\NP{Nucl. Phys.}  
\def\PLA{Phys. Lett.} 
\def\PLB{Phys. Lett.} 
\def\PL{Phys. Lett.} 
\def\PRL{Phys. Rev. Lett.}
\def\PRB{Phys. Rev. B}
\def\PRD{Phys. Rev. D}
\def\PR{Phys. Rev.}
\def\NC{Nuovo Cim.}
\def\ZPC{Z. Phys. C}
\def\SJNP{Sov. J. Nucl. Phys.}
\def\AnnP{Ann. Phys.\ (NY)}
\def\JETPL{JETP Lett.}
\def\LMP{Lett. Math. Phys.}
\def\CMP{Comm. Math. Phys.}
\def\PTP{Prog. Theor. Phys.}
\def\PPS{Proc. Phys. Soc.}
\def\RMP{Rev. Mod. Phys.}
\def\PRSA{Proc. Roy. Sci.} 
\def\PNAS{Proc. Nat. Acad. Sci.}
\def\PVAS{Proc. V.A. Steklov Inst. Math.}

\end{document}